\documentclass{elsart}
\journal{New Astronomy}
\begin{document}
\begin{frontmatter}
\title{The Magnetospheric Flare on Compact Magnetized Object ( Neutron Star
or White Dwarf )--\\ Model for Cosmological Gamma-Ray Burst
(GRB)}

\author[Fian]{Ya. N. Istomin\corauthref{cor}},
\corauth[cor]{Corresponding author.}
\ead{istomin@td.lpi.ac.ru}
\author[Asc]{B. V. Komberg}
\ead{bkomberg@asc.rssi.ru}

\address[Fian]{P.~N.~Lebedev Physical Institute,
Leninsky Prospect 53, Moscow, 119991 Russia}

\address[Asc]{Astro Space Center of the P.~N.~Lebedev Physical Institute,
Profsoyuznaya 84/32, Moscow, 119991 Russia}

\begin{abstract}

The SN explosion in the closed binary can give the
magnetospheric flare possessing the properties of GRB. The SN
shock, flowing around the magnetosphere
of a magnetized neutron star or a white dwarf, produces a narrow
magnetic tail $10^9 cm$ long, $10^8 cm$ wide and
a magnetic field of $10^6 Gauss$. Fast particles
( $\gamma\simeq 10^4$ ), generated in the tail by reconnection
processes, radiate gamma rays of the $100KeV$ - $1MeV$ energies. The duration
of radiation $T<1 sec$ corresponds to a short GRB. Apart, the powerful
shock can tear and accelerate part of the tail. That is the relativistic
( $\Gamma\simeq 10^4$ ), strongly magnetized jet, producing gamma
radiation and also X-ray and optic afterglow. That is a long ($T>10 sec$)
GRB. The duration of the afterglow is inversely proportional to the
photon energy and is several months for optic.

\end{abstract}


\end{frontmatter}

\section{Introduction}

Identification of two tens of the long GRB ( $T_{90}>1sec$ ) with
the extragalactic objects located at cosmological distances (
$z=0.1 - 4$ ) raises a question about the source of energy
release. For the spherical symmetry the energy yield is turned to
be as abnormally large as $10^{53-54} ergs$ only in the band of
thousands of $KeV$. Such energy release is comparable with that of a
supernova explosion ( SN ), for which $90\%$ of the
energy is escaped by neutrinos during the first several
seconds. To overcome the 'energy catastrophe' in the problem
of GRB a hypothesis was suggested about a strong
collimation of the gamma radiation in the narrow solid angle of
$\Delta\phi\simeq 1^\circ$. This assumption gives a possibility
to diminish the luminosity of GRB, needed for the explanation of
the observed gamma-ray flux, by a factor of $ (\Delta\phi/4)^2 $ which
for $\Delta\phi\simeq 1^\circ$ gives the gain of $10^5$. Such
directivity can be explained by the radiation of a narrow jet of
particles accelerated to relativistic energies. However, it
is not clear now how this jet is formed when two compact objects
merge or when one of them explodes. That is why there exists a
large amount of models suggested for GRB explanation, \cite{Djo}.
In last years the model of strong
anisotropic supernova explosion becomes very popular ( see, for example,
\cite{Pac} ),
which demands extreme conditions for its realization.

Our model suggests the mechanism of origin of a narrow beam of
gamma-rays when a shock of an SN explosion in
the binary system, containing a magnetized neutron star or a white
dwarf, is interacting with its magnetosphere. If the observer is placed in the
plane of the orbit of such a binary system and at the moment of
the origin of the magnetospheric tail the line of observation
coincides with the tail direction, then a GRB will be observed.
We give the estimations confirming that our model can
explain the general properties of GRB, both long ( $T>1sec$ ) --
cosmological, and short ( $T<1sec$ ) -- possibly local. For the
short GRB we consider the binary system in which one companion, as
previously, is a neutron star or a white dwarf and another is
a star of cataclysmic flare variable.

\section{Model's Parameters}

As an example we discuss a binary stellar system with a distance
between the companions of $ a\simeq 10^{13}cm$. The system
consists of a blue giant - supernova progenitor ( $ M=20M_\odot,
R=50R_\odot $ ) and a magnetized neutron star ( $ M=1.4M_\odot,
R=10^6 cm, B=10^{12} Gauss $ ) or a magnetized white dwarf ( $
M\simeq 1M_\odot, R=10^9 cm, B=10^9 Gauss $ ). According to the
review \cite{Im} the parameters of the shock are

kinetic energy $E_{kin}\simeq 10^{47} ergs$,

velocity $u\simeq 4\cdot 10^9 cm/sec$,

density $\varrho\simeq 10^{-8} g/cm^3\simeq 10^{16} protons/cm^3$,

width $h\simeq 10^9 cm$.

Under the impact of the shock on the dipole magnetic field of a
neutron star or a white dwarf a magnetospheric
tail forms, aligned along shock wave propagation. The
magnetospheric magnetic field is amplified and compressed by the
cumulative effect, forming the narrow tail. The dependencies of
the magnitude of the magnetic field in the tail $B_t$ and the
tail's diameter $d_t$ on the distance to the center of the
compact star $l$ follow from the condition of freezing of
the magnetic field into the shock plasma and the conservation of the
magnetic field flux
\begin{eqnarray}
B_t(l) = B^*\left (\frac{l}{r^\star}\right )^{1/2}; \nonumber \\
d_t(l) = r^*\left (\frac{l}{r^\star}\right )^{-1/4}.
\end{eqnarray}
Here $r^*$ is the Alfven radius
\begin{equation}
r^* = R\left (\frac{B^2}{4\pi\varrho u^2}\right )^{1/6},
\end{equation}
where the magnetic field pressure is equal to the pressure of the
shock matter; $B^*$ is the value of the magnetic field on the
Alfven radius
\begin{equation}
B^* = \left (4\pi\varrho u^2\right )^{1/2}.
\end{equation}
The length of the tail $L$ is of the order of shock's width $h$,
$$ L\simeq h. $$ Thus, a stretched ( $d_t << L$
), magnetized ( $B^\star \simeq 10^6 Gauss$ ), almost parallel
magnetospheric tail forms with a stored magnetic energy $ \epsilon_B =
\int {B_t}^2/8\pi\cdot \pi {d_t}^2 dl$,
\begin{equation}
\epsilon_B = \frac{1}{8} B^{1/2}{B^*}^{3/2}(Rh)^{3/2}.
\end{equation}
For neutron stars and for white dwarfs the values of $\epsilon_B$
are equal to
\begin{eqnarray}
{\epsilon_B}^{n.s.} = 4\cdot 10^{36}\left (\frac{B}{10^{12}G}\right )^{1/2}
\left (\frac{B^*}{10^6G}\right )^{3/2}\left (\frac{R^*}{10^6 cm}\right )
^{3/2}\left (\frac{h}{10^9 cm}\right )^{3/2} ergs, \nonumber \\
{\epsilon_B}^{w.d.} = 4\cdot 10^{39}\left (\frac{B}{10^{9}G}\right )^{1/2}
\left (\frac{B^*}{10^6G}\right )^{3/2}\left (\frac{R^*}{10^9 cm}\right )
^{3/2}\left (\frac{h}{10^9 cm}\right )^{3/2} ergs
\end{eqnarray}
respectively.
We see that for a not very compact pair ( $a\simeq 10^{13} cm$ )
a part of the energy caught from the total kinetic energy of the
shock of the SN explosion is $10^{-8}$ and $10^{-11}$ for the white
dwarfs and the neutron stars \footnote{We assume $E_{kin}\simeq
10^{47}ergs$, that is typical of a compact SN of Ib/c type. For
the Ia type $E_{kin}$ may $10^{3-4}$ times larger.}.
The energy caught increases as $a^{-3/2}$ for a closer
binary system.

For the narrow directivity of gamma radiation $\Delta\phi\simeq 10^{-4}$
from the relativistic particles ( $\gamma\simeq 10^4$ ), when the magnetospheric
tail is directed to the observer, it is natural to obtain the effective
power needed for the cosmological GRB. The rate of GRB of $300
year^{-1}$ for a total number of $10^{11}$ galaxies of a visible Universe
gives the rate per one galaxy of the order of $3\cdot 10^{-9}year^{-1}$.
If we assume
the rate of SN explosions in one young galaxy as $1SN year^{-1}$, then we
obtain that approximately $10^8$ SN explosions result in one GRB. This
corresponds to the directivity of gamma radiation of $\Delta\phi\simeq
10^{-4}$.

The lifetime of the tail is defined by the annihilation
of its magnetic field $\tau = L/v_a \equiv h/u$ ( $v_a$ is the
Alfven velocity ) and is equal to the time of its forming $h/u$.
It is $1 sec$. The non-stationarity and the reconnection of the
magnetic field in the tail ( note that the magnetic field is
antiparallel at the opposite sides of the tail ) lead to the
appearance of a strong electric field
\begin{equation}
E = \frac{d_t}{c\tau}B_t\simeq 10^{-2}B^*,
\end{equation}
accelerating  particles. Their mean Lorentz factor $\gamma$ can
be estimated from the equilibrium between the acceleration rate
and synchrotron losses of fast particles in the magnetic field
$B^*$,
\begin{equation}
\gamma = 10^4\left (\frac{r^*}{10^8 cm}\right )^{1/2}\left (\frac{u}{4\cdot
10^9 cm/sec}\right )^{1/2}\left (\frac{h}{10^9 cm}\right )^{-1/2}\left (
\frac{B^*}{10^6 G}\right )^{-1/2}.
\end{equation}
The characteristic value of the Lorentz factor of accelerated
electrons ( and positrons ) is $10^4$. In the magnetic field
$B^*\simeq 10^6 Gauss$ particles of such energies
radiate at the frequency $\nu\simeq\omega_c\gamma^2$, which
corresponds to the photon energy $500 KeV$, typical of a gamma
burst.

\section{Jet}

In some cases the magnetospheric tail produced by the interaction
of a SN shock with a magnetosphere of a magnetized companion
can be torn due to a global reconnecton of its magnetic field
lines ( tearing instability ). Such kind of phenomenon is observed
during intense chromospheric flares on the Sun. As a result,
part of the tail can be torn off and obtain high kinetic
energy. It will move with a Lorentz factor $\Gamma$, which is almost equal to
the factor $\gamma$, estimated early in (7), $\Gamma\simeq 10^4$.
Thus, a narrow, magnetized, relativistic jet is
moving towards the observer.

In the frame moving together with the jet, let the magnetic field
strength be $B$, the particle density be $n$, the mean Lorentz
factor of particles be $\gamma$. Due to synchrotron loses
the value of $\gamma$ will decrease. Inside the jet
microreconnection processes will support the equipartision between
the magnetic energy and the particle energy, $B^2 = 8\pi mc^2
\gamma n$. Besides, particles are frozen into the magnetic field,
and their density is proportional to the magnetic field strength,
$n=n_0 B/B_0$. The index "0" is related to the initial values of
the parameters. It follows that $$ B=8\pi mc^2n_0\gamma/B_0.
$$
During the adiabatic expansion of the jet's diameter with the
conservation of the magnetic flux, $d = d_0(B_0/B)^{1/2}$, the
quantities $n$ and $B$ will decrease in time. Because of the optical depth
with respect to the Thompson scattering is small $n_0 L\sigma_T
<<1$ for our parameters ($ n_0\simeq 10^{13} cm^{-3}, L\simeq 10^9
cm$ ) we can consider that the synchrotron radiation escapes
freely. So, the jet's cooling is associated with the synchrotron losses
\begin{equation}
mc^2\frac{d\gamma}{dt} = eEc -\frac{2e^4B^2\gamma^2}{3m^2c^3}.
\end{equation}
Here $E$ is the induced electric field arising due to the change of the
magnetic field, $E=-(r^*/c)dB/dt$. From (8) we obtain the equation describing
the evolution of the mean particle energy
\begin{equation}
\frac{d\gamma}{dt}\left [1+\frac{8\pi er^*n_0}{B_0}\right ] = -\frac{2e^4}
{3m^3 c^5}\left (\frac{8\pi mc^2n_0}{B_0}\right )^2 \gamma^4.
\end{equation}
The second term in the left-hand side of (9) describes the
betatron cooling of the particles with decreasing magnetic field
$$
\frac{8\pi r^*en_0}{B_0} =
\frac{2r^*{\omega_{p0}}^2}{c\omega_{c0}}.
$$
$\omega_{p0}$ is the
plasma frequency and $\omega_{c0}$ is the cyclotron frequency in
the jet at the initial time $t=0$. In our case for
${\omega_{p0}}^2 = 3\cdot 10^{22} sec^{-2}, \omega_{c0}=2\cdot
10^{13} sec^{-1}, r^* = 10^8 cm$ the value of
$2r^*{\omega_{p0}}^2/c\omega_{c0}>>1$, and we can write the
solution of equation (9) as
\begin{equation}
\gamma^{-3} - \gamma_0^{-3} = \frac{t}{\tau_0},
\end{equation}
where the time $\tau_0 = r^*\omega_{c0}/4\omega_{p0}^2 r_e\simeq
2\cdot 10^4 years$ ( $r_e$ is the classic electron radius ). For
not very small times in the frame of the jet, $t>\tau_0
\gamma_0^{-3}$, we have
\begin{eqnarray}
\gamma = \left (\frac{t}{\tau_0}\right )^{-1/3}, \nonumber \\
B\propto\left (\frac{t}{\tau_0}\right )^{-1/3}.
\end{eqnarray}
The characteristic frequency $\omega = \omega_c \gamma^2$, at
which relativistic particles radiate synchrotron photons,
will change in time $$ \omega =
\frac{2{\omega_{p0}}^2}{\omega_{c0}}\left (\frac{t}{\tau_0} \right
)^{-1}\propto\frac{1}{t}. $$ For such a dependence $\omega(t)$
the observed frequency $\omega^{'} = \Gamma\omega$ and the time in
the frame of observer $t^{'}=t/\Gamma$ are related as
$$ t^{'} =
\frac{2\tau_0{\omega_{p0}}^2}{\omega_{c0}\omega^{'}}\propto\frac{1}
{\omega^{'}}. $$ We see that the time of glowing at a given
frequency in the observer frame is $$ t^{'}= \frac{2\cdot 10^5
eV}{\hbar\omega^{'} eV} sec. $$ For the photon energy
$\hbar\omega^{'}\simeq 0.3 MeV$, which is typical of a GRB, the time of
glowing is $10 sec$. This time just corresponds to the maximum of
the distribution of long GRB over their duration ( at the level 90\%
from the peak luminosity - $T_{90}$ ). The observations showed
that it is long GRB ( $T_{90} > 10 sec$ ) that are sometimes accompanied
the afterglow in X-ray and/or optic bands. According to our model
the duration of the X-ray afterglow ( $\hbar\omega^{'}\simeq 100 eV$ ) is
several days, and that of the optic afterglow ( $\hbar\omega^{'}\simeq 1
eV$ ) is several months. These estimations do not
contradict the observations of GRB.

From our model we obtain also the estimations for the rate of the change
of the afterglow intensity, $I(\omega^{'})\propto{\omega^{'}}^{4/3}\propto
(t^{'})^{-1.33}$. The observations show ( see, for example \cite{Djo}
, \cite{Sim} ) the
dependence $I\propto (t^{'})^{-1.2}$, which does not contradict our estimation.

Probably, if the energy of the shock from a SN is not enough to
tear off the magnetospheric tail of magnetized component, then there
will not be any jet and consequently  no afterglow will be observed.
This situation is more typical of a not energetic
GRB and they will be short ( $T_{90} <1 sec$ ).

\end{document}